\documentclass{article}
\usepackage[margin=1.5in]{geometry}

\usepackage{apacite}
\usepackage{authblk}
\usepackage{amsmath,amssymb}
\usepackage{comment}
\usepackage{graphicx}
\usepackage{makecell,multicol,multirow}
\usepackage{subfig}
\usepackage{url}
\usepackage[utf8x]{inputenc}

\usepackage[most]{tcolorbox}

\title{Social Bots: Detection and Challenges}
\author[1,2]{Kai-Cheng Yang (0000-0003-4627-9273)}
\author[3]{Onur Varol (0000-0002-3994-6106)}
\author[4]{Alexander C. Nwala (0000-0003-3408-791X)}
\author[1]{Mohsen Sayyadiharikandeh}
\author[5]{Emilio Ferrara (0000-0002-1942-2831)}
\author[1]{Alessandro Flammini (0000-0003-1670-9156)}
\author[1]{Filippo Menczer (0000-0003-4384-2876)}
\affil[1]{Luddy School of Informatics, Computing, and Engineering, Indiana University Bloomington, USA}
\affil[2]{Network Science Institute, Northeastern University, USA}
\affil[3]{Faculty of Engineering and Natural Sciences, Sabanci University, Turkey}
\affil[4]{Department of Applied Science, William \& Mary, USA}
\affil[5]{Thomas Lord Department of Computer Science, University of Southern California, USA}

\newcolumntype{.}{!{\vrule width 2pt}}
\newcolumntype{'}{!{\vrule width 1.5pt}}

\begin{document}
\maketitle

\begin{tcolorbox}[colback=black!3,colframe=white!45!black]
This is a draft of the chapter. The final version will be available in the Handbook of Computational Social Science edited by Taha Yasseri, forthcoming 2024, Edward Elgar Publishing Ltd. The material cannot be used for any other purpose without further permission of the publisher and is for private use only. Please cite as: Yang, K.-C., Varol, O., Nwala, A. C.,  Sayyadiharikandeh, M., Ferrara, E., Flammini, A., and Menczer, F. (2024). Social Bots: Detection and Challenges. In: Yasseri, T. (Ed.), \textit{Handbook of Computational Social Science}. Edward Elgar Publishing Ltd.
\end{tcolorbox}

\abstract{
While social media are a key source of data for computational social science, their ease of manipulation by malicious actors threatens the integrity of online information exchanges and their analysis. 
In this Chapter, we focus on malicious social bots, a prominent vehicle for such manipulation. 
We start by discussing recent studies about the presence and actions of social bots in various online discussions to show their real-world implications and the need for detection methods.
Then we discuss the challenges of bot detection methods and use Botometer, a publicly available bot detection tool, as a case study to describe recent developments in this area.
We close with a practical guide on how to handle social bots in social media research.
}

\begin{center}
\textbf{Keywords}: social bots, Twitter, bot detection, astroturfing, online manipulation, information spreading
\newline
\end{center}

\section{Introduction}

Social media are an important source of data for computational social science studies~\cite{doi:10.1126/science.aaz8170}.
On the one hand, with more people joining the online community, the virtual and the real world have become more intertwined than ever, producing new phenomena and research questions.
On the other hand, massive digital traces of user activity make it possible to characterize these phenomena and address these research questions.
Many aspects of computational social science, such as those covered by this Handbook, can be affected by malicious actors that attempt to disrupt healthy online communication. 
In this Chapter, we focus on malicious social bots on Twitter (rebranded to X in 2023), a prominent type of such actors.

Social bots are social media accounts controlled in part by algorithms that can automatically post content and interact with other accounts~\cite{ferrara2016rise}.
A 2017 study estimated that 9--15\% of active Twitter accounts were bots~\cite{varol2017online}.
Although social media platforms have strengthened their efforts to contain malicious actors in recent years, bots remain prevalent and their tactics to evade detection continue to evolve~\cite{yang_arming_2019}. This has two implications for computational social science practitioners.
First, characterizing the behavior and assessing the impact of social bots remain relevant topics of investigation~\cite{rahwan_machine_2019}.
Second, researchers need to properly account for how bots may distort data and analyses~\cite{jamison_malicious_2019,ledford_social_2020}. 
Indeed, the number of bot-related studies has grown rapidly in recent years (see Figure~\ref{fig:publication_ts}; methods below) in parallel with the quest for accessible and reliable tools that can detect social bots.
The main goal of the present chapter is to provide readers with sufficient knowledge to independently conduct research on social bots or to identify and possibly eliminate bot activity from their datasets.

The Chapter is organized as follows.
First, we offer a review of the literature related to social bots.
Then we discuss recent progress and challenges in bot detection methods. 
Finally, we provide a practical guide to performing bot detection for research.

\section{State-of-the-art}

\subsection{Scientometric analysis}

To illustrate the research landscape related to social bots, we conducted a scientometric analysis.
We collected publications records using the Dimensions API,\footnote{\url{www.dimensions.ai}} which provides details about publications such as venue, publication date, title, authors, and so on.

\begin{figure}
  \centering
  \includegraphics[width=0.49\linewidth]{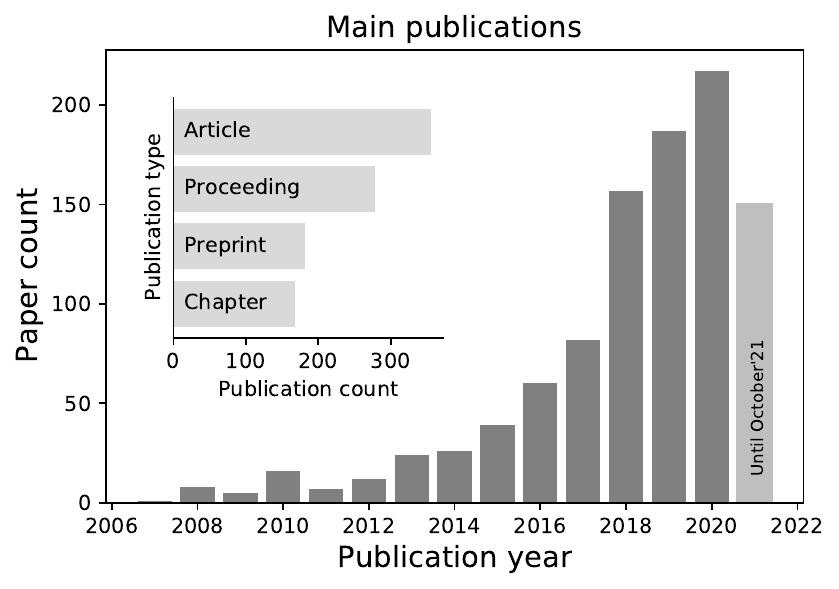} 
  \includegraphics[width=0.49\linewidth]{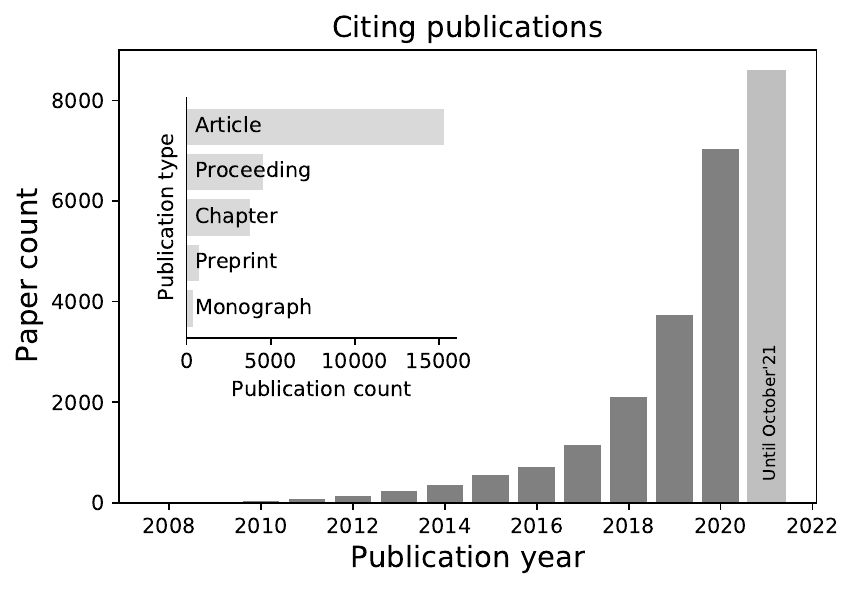}
  \caption{
  Numbers of publications regarding social bots (left) and publications citing them (right) over time.
  Publication types are shown in the insets.
  }
  \label{fig:publication_ts}
\end{figure}

The Dimensions system computationally extracts keywords and assigns relevance scores for each paper.
We identified relevant papers by filtering based on their relevance scores. The list of keywords used for filtering consisted of phrases such as ``bot detection,'' ``social bots,'' and ``political bots.''  
Our collection resulted in 911 publications authored by 1,618 distinct authors.
We also gather references and citations for each paper in the collection.
The numbers of relevant publications and publications citing them over time are shown in Figure~\ref{fig:publication_ts} together with publication types.
We observe that social bot research has attracted significant attention in recent years, leading to more than 200 publications and over 7,000 citations in 2020 alone.

\begin{figure}
  \centering
  \includegraphics[width=0.68\linewidth]{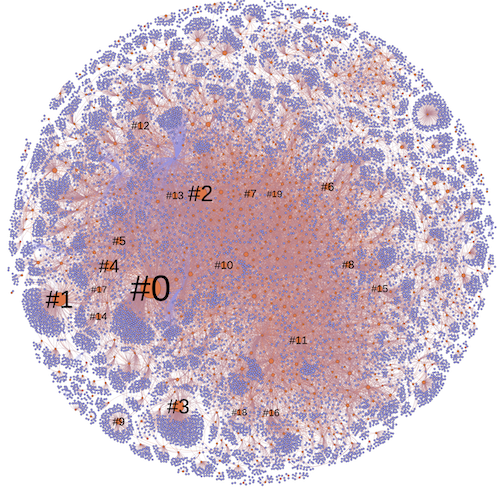} 
  \includegraphics[width=0.73\linewidth]{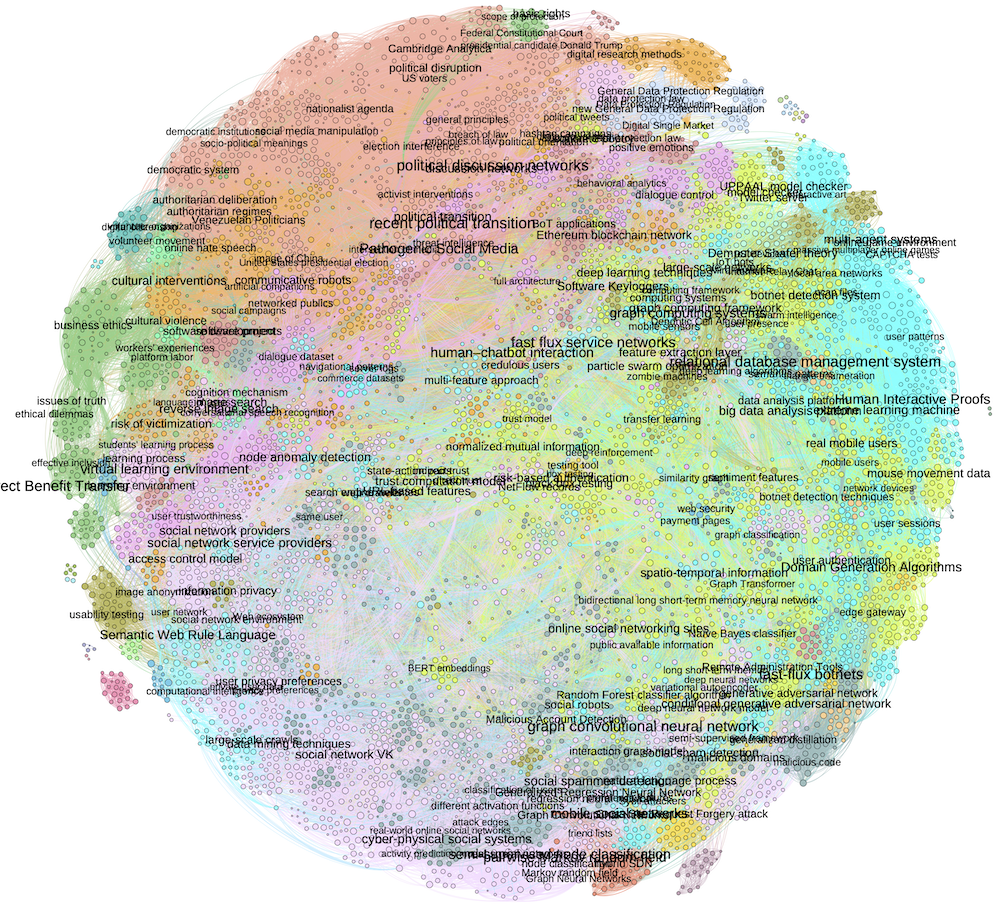}
  \caption{
  Network analyses of publications about social bots. 
  Top: Citation network, with papers about social bots and citing publications represented as orange nodes and purple colors, respectively. Top-cited social bot papers are referred to with numbers and listed in Table~\ref{tab:papers}. Bottom: Topical network, with node colors corresponding to communities of papers identified by modularity maximization.
  }
  \label{fig:publication_networks}
\end{figure}

To identify the central studies in this area, we mapped the citation network among the social bot papers, additionally including all other publications citing those papers.
The extended collection consists of 24,717 papers.
Figure~\ref{fig:publication_networks} visualizes the citation network.
The nodes corresponding to top-cited articles are larger, and the top-ranked papers are labeled and listed in Table~\ref{tab:papers}. 

\begin{table}[t]
    \caption{Paper numbers refer to Fig.}
    \centering
    \resizebox{\columnwidth}{!}{
    \begin{tabular}{c|p{8cm}|p{5cm}}
        \hline
        Paper \# & Title & Reference \\
        \hline \hline
        \texttt{\#0} & The rise of social bots & \citeA{ferrara2016rise}\\
        \hline
        \texttt{\#1} & Weaponized health communication: Twitter bots and Russian trolls amplify the vaccine debate & \citeA{broniatowski2018weaponized}\\
        \hline
        \texttt{\#2} & Botornot: A system to evaluate social bots  & \citeA{davis2016botornot}\\
        \hline
        \texttt{\#3} & Exposure to opposing views on social media can increase political polarization & \citeA{bail2018exposure}\\
        \hline
        \texttt{\#4} & The spread of low-credibility content by social bots & \citeA{shao_spread_2018}\\
        \hline
        \texttt{\#5} & Bots increase exposure to negative and inflammatory content in online social systems & \citeA{stella2018bots}\\
        \hline
        \texttt{\#6} & Detecting spam bots in online social networking sites: A machine learning approach & \citeA{wang2010detecting}\\
        \hline
        \texttt{\#7} & Dissecting a social botnet: Growth, content and influence in Twitter & \citeA{abokhodair2015dissecting}\\
        \hline
        \texttt{\#8} & Tweets as impact indicators: Examining the implications of automated ``bot'' accounts on Twitter & \citeA{haustein2016tweets}\\
        \hline
        \texttt{\#9} & Measuring price discrimination and steering on e-commerce web sites & \citeA{hannak2014measuring}\\
        \hline
    \end{tabular}
    }
    \label{tab:papers}
\end{table}

Social bot research contributes to other areas as well. Its widespread influence is illustrated in the topical network in Figure~\ref{fig:publication_networks}.
Different publication communities were identified by the modularity maximization approach implemented in Gephi~\cite{bastian2009gephi}.
Analysis of the categories assigned to the citing publications by the Dimensions API reveals the fields that benefit from research on social bots, and ``Computer Science and Informatics,'' ``Business and Management Studies,'' and ``Communication, Cultural and Media Studies'' are the top three.
Social science disciplines such as ``Politics and International Studies,'' ``Sociology,'' and ``Law'' are among the top 15.

\subsection{Bot behaviors and activities}

Several of the studies in the above analysis focus on identifying the presence of malicious bots in different contexts and assessing their impact.
Bots are involved in all types of online discussions, especially controversial ones.
The most popular topic is political elections: studies report social bots activity in the context of U.S. elections~\cite{shao_spread_2018,gorodnichenko_social_2021,bessi_social_2016,ferrara_characterizing_2020}, French elections~\cite{ferrara_disinformation_2017}, the Brexit referendum~\cite{bastos_public_2018,bastos_brexit_2019,gorodnichenko_social_2021,duh_collective_2018}, German elections~\cite{keller_social_2019}, and the 2017 Catalan independence referendum~\cite{stella2018bots}.
Public health is another area in which concerning activity by malicious bots has been reported~\cite{jamison_malicious_2019}: bots actively participated in the debates regarding vaccines~\cite{broniatowski_weaponized_2018,yuan_examining_2019}, the COVID-19 pandemic~\cite{ferrara_what_2020,shi_social_2020,uyheng_bots_2020,yang2020prevalence}, and cannabis~\cite{allem_cannabis_2020}.
Other research uncovered the presence of social bots in discussions about climate change~\cite{marlow_bots_2021,chen_social_2021}, cryptocurrency~\cite{nizzoli_charting_2020}, and the stock market~\cite{cresci_cashtag_2019,fan_social_2020}.

Malicious social bots exhibit a variety of behavioral patterns. 
Some simply generate a large volume of posts to amplify certain narratives~\cite{marlow_bots_2021,keller2020political} or to manipulate the price of stocks~\cite{cresci_cashtag_2019,fan_social_2020} and cryptocurrencies~\cite{nizzoli_charting_2020}.
Others disseminate low-credibility information strategically by getting involved in the early stage of the spreading process and targeting popular users through mentions and replies~\cite{shao_spread_2018}.
Some bots are used as fake followers to inflate the popularity of other accounts~\cite{bilton_social_2014,confessore_follower_2018,zouzou2023unsupervised}. Analysis of anomalous followers unveiled the usage of bots to promote unverified Twitter accounts who self-identify as journalists~\cite{varol2020journalists}.
In terms of content, malicious bots have been found to engage other accounts with negative or inflammatory language~\cite{stella2018bots} or hate speech~\cite{albadi_hateful_2019,uyheng_bots_2020}.
In some cases, bots contribute to dense social networks to boost popularity metrics and amplify the diffusion of each other's messages~\cite{caldarelli_role_2020,torres2020trains,chen_neutral_2021}.

Although this Chapter focuses on malicious bots, we stress that not all bots are designed with malicious intent. 
Social bots can contribute to public good by automatically sharing news and disaster information~\cite{lokot_news_2016}, disseminating research papers~\cite{haustein_tweets_2016}, delivering targeted health interventions~\cite{deb_social_2018}, organizing social movements~\cite{flores-saviaga_leadwise_2016,savage_botivist_2016}, and so on.
However, good intentions do not always translate into good outcomes~\cite{tsvetkova_even_2017,rahwan_machine_2019}.
For example, recent work by \citeA{chen_neutral_2021} shows that non-intelligent bots designed to be neutral start to generate low-credibility information after interacting with other accounts for several months. This suggests that bots can unintentionally act as amplifiers of malicious actors.

Bots are also used as tools to explore research questions in computational social science that are not directly concerned with the bots themselves.
For example, some studies explore the factors affecting the human perception of social bots~\cite{yan_asymmetrical_2021,wischnewski_disagree_2021}. 
Researchers also used bots as instruments to reveal the mechanism of online information spread~\cite{monsted_evidence_2017}, to understand why people share misinformation~\cite{pennycook2021shifting}, and to probe political bias in social media~\cite{chen_neutral_2021}.
In other studies, researchers use bot detection tools to focus their analyses on online human activity~\cite{grinberg_fake_2019,bovet_influence_2019}.

Most of the studies mentioned above depend on the ability to identify social bots.
In the next section, we summarize recent developments in the area of bot detection methods on Twitter.

\section{Bot detection methods and challenges}

Researchers have identified bot-like accounts through methods such as heuristics \linebreak \cite{gorodnichenko_social_2021} and human annotation~\cite{gilani2017bots}.
Heuristic rules such as identifying accounts posting content at an unusually high frequency are easy to implement, but it is nontrivial to build a rule-based system that can capture the complex behavioral patterns of different bot classes.
Human annotation can provide reliable results, but it becomes infeasible when dealing with large-scale data.
More commonly, researchers resort to bot detection tools that employ machine learning models.
Such tools are good at characterizing the complex behaviors of different social media accounts and can scale up to datasets with millions of entities.
Readers can refer to excellent reviews of various bot detection tools proposed in recent years by \citeA{orabi_detection_2020} and \citeA{cresci2020decade}. 

\begin{figure}
  \centering
  \includegraphics[width=\linewidth]{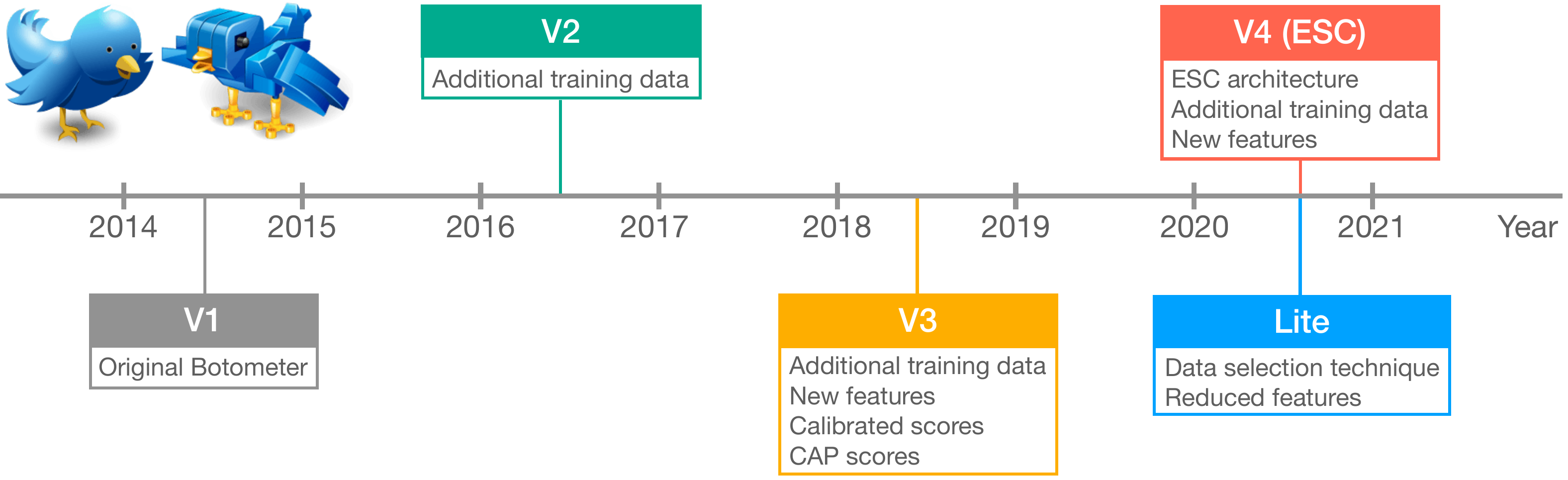}
  \caption{
  Timeline of Botometer.
  }
  \label{fig:botometer_timeline}
\end{figure}

Let us focus on Botometer,\footnote{\url{botometer.org}} a supervised machine learning tool for bot detection, to provide an in-depth examination of the challenges and evolution of bot detection methods. This choice is motivated by three factors. 
First, unlike most other bot detection tools proposed in the literature, Botometer is publicly available through a web interface and API (Application Programming Interface) endpoints.
Anyone with a Twitter account can use the web version for free; researchers with Twitter developer accounts can use the API endpoints to analyze large-scale datasets.
Second, owing to its public availability, Botometer has been used in many studies~\cite[among others]{shao_spread_2018,bovet_influence_2019,grinberg_fake_2019}. 
Finally, and most importantly, the evolution of Botometer reflects the general progress of bot detection methods over a long time: versions of the tool have been running for over seven years (see the timeline in Figure~\ref{fig:botometer_timeline}).

The early versions of Botometer (V1 and V2) had a relatively simple architecture. 
They extracted over 1,000 features from user profiles, content, temporal patterns, and social networks, and fed these features to a Random Forest classifier for evaluation~\cite{davis2016botornot,varol2017online,varol2018feature}.
The output of the classifier was returned to users directly as the bot score, where a higher score indicated that the account was more bot-like.
Most of the supervised machine learning models for bot detection in the literature have similar architectures, though they may differ in training datasets, feature engineering, and classification algorithms.
Although the early versions of Botometer had decent accuracy (90\%-94\% AUC score of ROC), several challenges emerged over time.

\subsection{Score misinterpretation}

One challenge of bot detection is how to correctly interpret the results of the classifier.
Botometer, at its core, uses a Random Forest classifier.
The result of the evaluation, the bot score, is a number in the unit interval representing the fraction of trees that classify the target account as a bot. 
By offering Botometer to a wide, typically non-technical user base, we learned that one cannot expect users to understand exactly what the bot score means.
It was very common for users to interpret the score as a probability: a score of x\% meaning that x\% of the accounts with that score are automated, which is \emph{not} what the score defined above means.

Addressing this issue required a change in the user interface: in later versions of Botometer, the score is reported on a scale between zero and five rather than as a fraction or percentage.

\subsection{Dependency on platform API}

A common issue that is often overlooked by researchers when proposing new bot detection methods is the dependency on the platform API.
All bot detection methods must fetch data from social media platforms before they can perform the evaluation.

Data accessibility is an obstacle in many cases.
For example, Facebook's unwillingness to share individual account data with researchers makes studying bots on that platform very hard.
Twitter used to have a more open data-sharing policy so that bot detection methods could be built on public data they make available via their APIs, although even more revealing information might be non-public.
Researchers recently made several requests to social media platforms aimed at improving data accessibility to tackle challenges such as the spread of misinformation and the detection of malicious bots~\cite{pasquetto2020tackling}.
Alternatively, researchers can organize to collect massive-scale data by sharing their application keys to capture a complete picture of Twitter's public stream~\cite{pfeffer2023just}.

After Elon Musk acquired Twitter and rebranded it as X in 2023, he made several changes to the platform, and one of the most critical developments for the researchers was the removal of free API access. 
This change has led to the cancellation or suspension of many ongoing research projects relying on Twitter data.\footnote{\url{reuters.com/technology/elon-musks-x-restructuring-curtails-disinformation-research-spurs-legal-fears-2023-11-06}}
These projects include studies on spammer behaviors and social bots, topics of considerable importance in the negotiations between Musk and the former Twitter management before the acquisition~\cite{varol2023should}.
The recent enactment of the Digital Services Act by the European Union, which mandates that large online platforms, including Google, Meta, and Twitter, provide data access to researchers, offers a potential remedy.
Nonetheless, the actual feasibility of researchers obtaining the necessary data for their studies under this new regulation remains uncertain.

Another obstacle comes from scalability.
The speed at which a method can process a group of Twitter accounts depends on the rate limits of Twitter API endpoints.
For instance, information about the social network structure of accounts can be very informative in bot detection, but fetching this information is very time-consuming.
As a result, many bot detection methods cannot process large datasets.

There are issues of robustness against changes in platform API policies and specifications. 
For example, due to privacy considerations, Twitter removed the field \texttt{geo\_enabled} --- which was an important feature for bot detection --- from their API endpoints by setting it to ``false'' as the default in 2019. 
The transition from V1 to V2 of the Twitter API also strongly affects bot detection methods.
Changes to platform APIs often require corresponding adjustments to the models.

\subsection{Generalizability}

\begin{table}
\caption{
Annotated datasets of human and bot accounts used to train different versions of Botometer.
For BotometerLite, the datasets with \checkmark* are used to select the training datasets; more details are found in the main text.}
\centering
\resizebox{\columnwidth}{!}{
\begin{tabular}{lp{3cm}p{3cm}lllll}
   \hline
   Dataset & Annotation method & Ref. & V1 & V2 & V3 & V4 & Lite\\
   \hline
   \texttt{caverlee}  & Honeypot + verified & \citeNP{lee2011seven} & \checkmark & \checkmark & \checkmark & & \\
   \texttt{varol-icwsm} & Human annotation & \citeNP{varol2017online} & & \checkmark & \checkmark & \checkmark & \checkmark \\
   \texttt{cresci-17} & Various methods & \citeNP{cresci2017paradigm} & & & \checkmark & \checkmark & \checkmark \\
   \texttt{pronbots} & Spam bots & \citeNP{yang_arming_2019} & & & \checkmark & \checkmark &\\
   \texttt{celebrity} & Celebrity accounts & \citeNP{yang_arming_2019} & & & \checkmark & \checkmark & \checkmark \\
   \texttt{vendor-purchased} & Fake followers &\citeNP{yang_arming_2019} & & & \checkmark & \checkmark &\\
   \texttt{botometer-feedback} & Human annotation & \citeNP{yang_arming_2019} & & & \checkmark & \checkmark & \checkmark \\
   \texttt{political-bots} & Human annotation & \citeNP{yang_arming_2019} & & & \checkmark & \checkmark & \checkmark \\
   \texttt{gilani-17} & Human annotation  & \citeNP{gilani2017bots} & & & & \checkmark & \checkmark*\\
   \texttt{cresci-rtbust} & Human annotation &\citeNP{mazza_rtbust_2019} & & & & \checkmark & \checkmark*\\
   \texttt{cresci-stock} & Sign of coordination & \citeNP{cresci_fake_2018} & & & & \checkmark &\\
   \texttt{verified} & Human annotation  & \citeNP{yang_scalable_2020} & & & & & \checkmark*\\
   \texttt{botwiki} & Self-declared  & \citeNP{yang_scalable_2020} & & & & \checkmark & \checkmark*\\
   \texttt{midterm-2018} & Human annotation & \citeNP{yang_scalable_2020} & & & & \checkmark & \checkmark*\\
   \texttt{astroturf} & Human annotation & \citeNP{sayyadiharikandeh2020detection} & & & & \checkmark &\\
   \texttt{kaiser} & Politicians + new bots  & \citeNP{harvardDataset} & & & & \checkmark &\\
   \hline
\end{tabular}
}
\label{table:dataset} 
\end{table}

The accuracy of supervised machine learning models, like Botometer, relies heavily on the representativeness of their training data.
As few collections of labeled Twitter accounts were available, the early versions of Botometer were trained on just a few datasets (see Table~\ref{table:dataset}).

Generalizability is a critical challenge: accuracy declines when dealing with accounts that are significantly different from those in the training datasets. 
Accounts might come from a different context, use languages other than English~\cite{rauchfleisch_false_2020,martini_bot_2021}, or display novel behavioral patterns~\cite{cresci2017paradigm,yang_arming_2019, Dimitriadis_multiclass_2021}.

\begin{figure}
  \centering
  \includegraphics[width=0.8\linewidth]{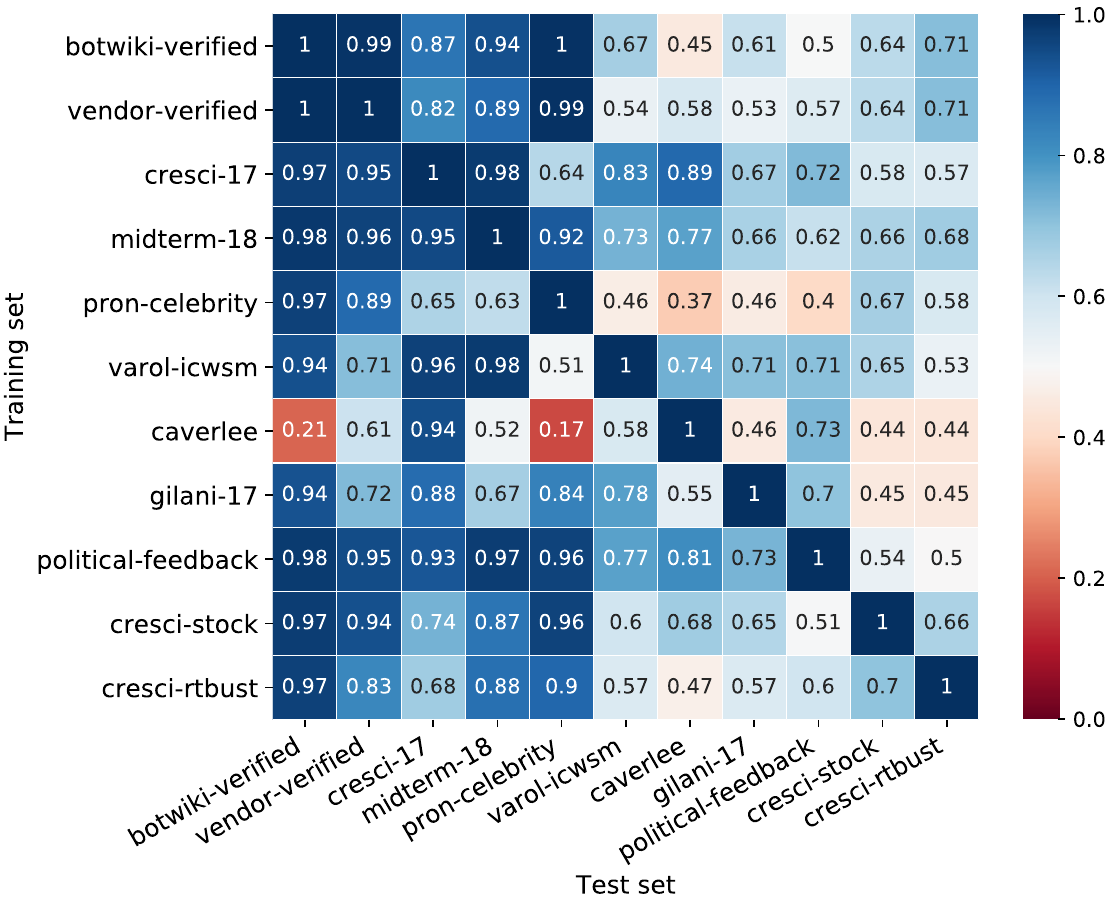}
  \caption{
  AUC scores of Random Forest classifiers trained on one dataset (row) and tested on another (column). Details about the datasets can be found in Table~\ref{table:dataset}. Source: \protect\citeA{yang_scalable_2020}.}
  \label{fig:generalizability}
\end{figure}

To better illustrate this issue, Figure~\ref{fig:generalizability} shows the results of an experiment using the datasets in Table~\ref{table:dataset}.
Different Random Forest classifiers are trained on single specific datasets (rows) and tested on different ones (columns).
Accuracy is measured using the Area Under the receiver operating characteristic Curve (AUC) score.
A higher AUC score indicates that the classifier can better discriminate between bot and human accounts in the test dataset.
In many of the off-diagonal cells, where the training and test sets differ, the AUC scores are relatively small. This suggests that the accuracy of a classifier is low when facing accounts different from those used to train it.

The generalizability challenge of supervised machine learning approaches such as that implemented in Botometer is largely due to the scarcity of representative datasets. 
Typical solutions involve adding more diverse samples to avoid overfitting the learning models to the training data.
Procuring such datasets, given the continuous appearance of new classes of bots, is overwhelmingly expensive as it typically requires human annotation. 
An alternative is the automatic search for accounts that act in coordination, 
an approach that has recently drawn some interest~\cite{pacheco2020uncovering,nizzoli2021coordinated,sharma2021identifying}.
Coordinated accounts often appear to be normal when inspected individually but participate in orchestrated actions to fulfill the agenda of the actors in control. 
Not all coordinated accounts are automated, but social bots provide a practical and low-cost means to deploy coordinated campaigns. 
Detecting coordinated accounts requires a group-level unsupervised approach, which typically involves defining a similarity metric among accounts and then clustering them~\cite{cresci2020decade}. 
While Botometer was not designed to identify new classes of bots via their coordinated behavior, it can be trained using datasets obtained in this way (see Table~\ref{table:dataset}). 
Section~\ref{v3v4} discusses other ways to address the generalizability issue.

\section{Recent developments}
\label{v3v4}

As discussed above, for a supervised bot detection tool to remain relevant, researchers need to constantly re-adjust its training datasets, as well as introduce smarter and/or more robust machine learning architectures. 
Let us now examine how these challenges are addressed by later versions of Botometer (cf. Figure~\ref{fig:botometer_timeline}).

\subsection{Botometer V3}

Built upon the early versions of Botometer, V3 included new training datasets (see Table~\ref{table:dataset}) and new features~\cite{yang_arming_2019}.
The new training datasets consisted of novel types of social bots.
The new features allowed the machine learning model to characterize accounts in new dimensions in the feature space, increasing its power to distinguish different types of accounts.
As a result, Botometer V3 achieved a higher accuracy and could identify new types of bots.

Botometer V3 also started to return calibrated scores in place of the raw scores returned by the Random Forest classifiers.
Consider two Twitter accounts $A$ and $B$ such that a classifier returns bot scores $C(A) = x$ and $C(B) = y$ with $x < y$ ($x, y \in [0, 1]$).
Here, one can say that account $A$ is less likely to be a bot than account $B$, but one cannot say that account $A$ is a bot with probability $x$ or that it is $(x \times 100)\%$ bot.
It is not straightforward for the typical users of Botometer to understand this difference, so the newly calibrated scores were included to bridge the gap between the classifier output and user expectations.

Calibrated scores come from a calibrated classifier whose output can be interpreted as a probability, i.e., $C'(A) = x$ means that the classifier estimates the probability that account $A$ is a bot as $x$.
To achieve this, Botometer V3 employed Platt's scaling~\cite{niculescu2005predicting}, a logistic regression model trained on classifier outputs.
The mapping shifts scores within the unit interval but preserves order, therefore having no impact on the model accuracy.

As mentioned earlier, the percentage of bot accounts was estimated to be around 9--15\%~\cite{varol2017online}, meaning that a randomly selected account has a low probability of being automated.
The calibrated score cannot reflect this fact since it is a likelihood estimation based on the training datasets.
We, therefore, introduced the Complete Automation Probability (CAP) score, which uses Bayesian posterior probabilities to overcome this problem. 
The connection between likelihood and background probability is formalized by Bayes' theorem. 
Denote $P(\textrm{Bot} \mid S)$ as the desired conditional probability that an account is a bot given its bot score $S$ (CAP).
Applying Bayes' rule allows us to rewrite this \textit{posterior} probability as:
\begin{equation}
   P(\textrm{Bot} \mid S) = P(\textrm{Bot})
  \frac{P(S \mid \textrm{Bot})}{P(S)},
  \label{eq:bayes}
\end{equation}
where the \textit{prior} probability $P(\textrm{Bot})$ is the background probability that any randomly chosen account is a bot.
The $\frac{P(S \mid \textrm{\scriptsize Bot})}{P(S)}$ term is called the \textit{evidence}.
It compares the \textit{likelihood} that a bot has score $S$, $P(S \mid \textrm{Bot})$, with the probability that any account has that score.

To calculate the posterior probability, the denominator in the evidence term of Eq.~\ref{eq:bayes} can be expanded as follows:
\begin{equation}
    \begin{split}
    P(S) & = P(S \mid \textrm{Bot}) \; P(\textrm{Bot})
    + P(S \mid \textrm{Human}) \; P(\textrm{Human}) \\
         & = P(S \mid \textrm{Bot}) \; P(\textrm{Bot})
    + P(S \mid \textrm{Human}) (1 - P(\textrm{Bot})).
    \end{split}
\end{equation}
The task is then to obtain distributions for the likelihoods $P(S \mid \textrm{Bot})$ and $P(S \mid \textrm{Human})$.
The training data provides empirical distributions of scores for both humans and bots.
We fitted these distributions against the family of Bernstein polynomials~\cite{babu2002} to create probability density functions that are likely to produce the empirical distributions. 
For the prior term $P(\textrm{Bot})$, which represents the background probability of a randomly chosen account being a bot, Botometer V3 used $P(\textrm{Bot})=0.15$, corresponding to the estimate mentioned above.

\subsection{Botometer V4}

Botometer V4 was a major upgrade.
Similarly to V3, V4 used new training datasets (see Table~\ref{table:dataset}) and features.
More importantly, Botometer V4 adopted a new architecture to improve its ability to generalize to accounts in datasets not used in training~\cite{sayyadiharikandeh2020detection}.

\begin{figure}
    \centering
    \includegraphics[width=0.8\linewidth]{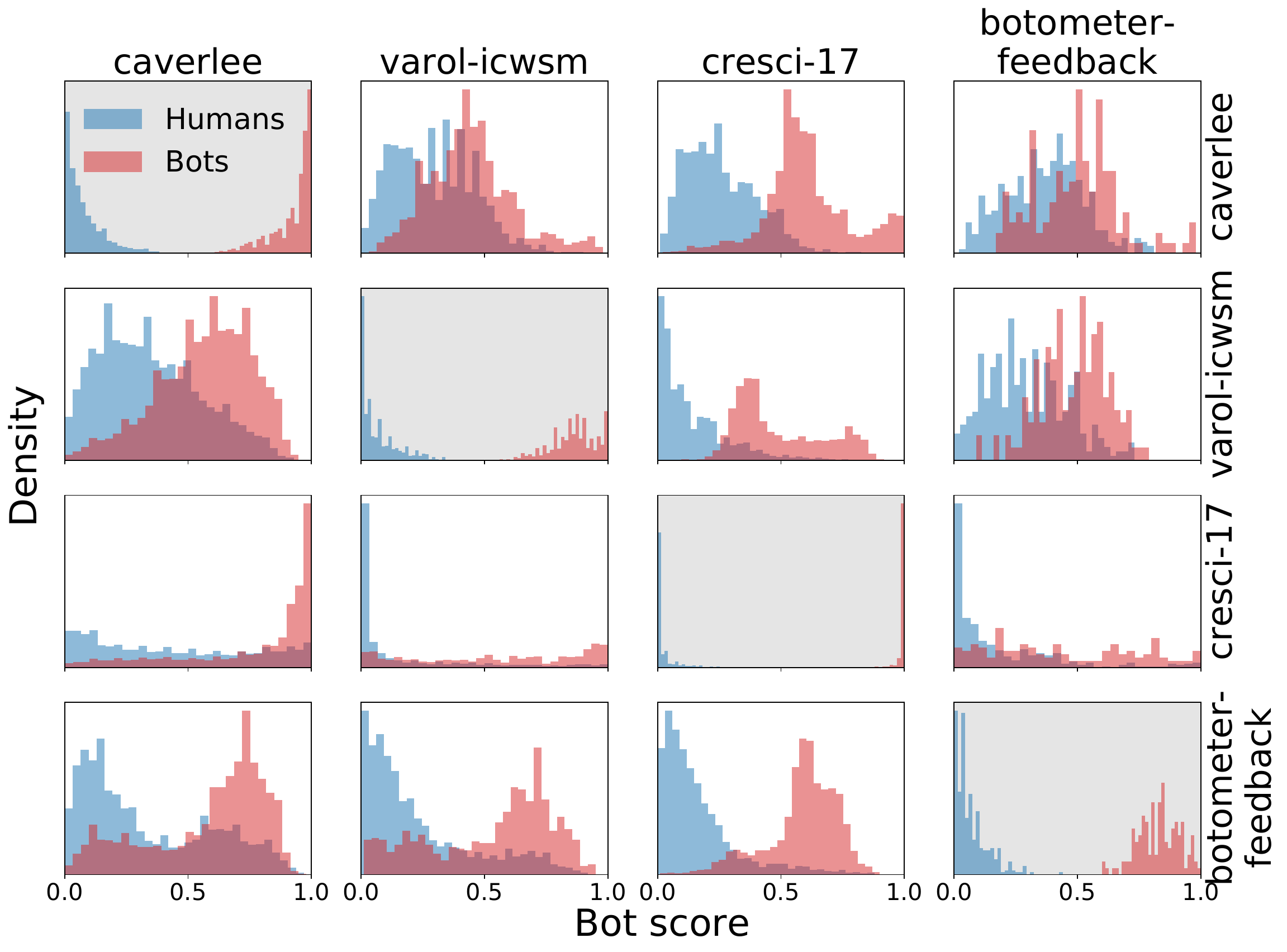}
    \caption{
    Bot score distributions for human (blue) and bot (red) accounts for Random Forest classifiers trained on the row datasets and tested on the column datasets.
    Source: \protect\citeA{sayyadiharikandeh2020detection}.
    }
    \label{fig:densityPlotsMatrix}
\end{figure}

The new architecture was inspired by two observations.
First, human accounts tend to have more homogeneous behaviors, while different types of social bots often show different characteristics. 
This is illustrated in Figure~\ref{fig:densityPlotsMatrix}, where classifiers trained on one dataset were tested on a different dataset.
The resulting bot score distributions are left-skewed for human accounts but not necessarily right-skewed for bot accounts.

\begin{table}
  \caption{Most informative features for different bot classes in the \texttt{cresci-17} dataset. Source:  \protect\citeA{sayyadiharikandeh2020detection}.}
  \label{tab:botclassFeatures}
  \footnotesize
  \centering
  \resizebox{\columnwidth}{!}{
  \begin{tabular}{llll}
  \hline
  Rank & Traditional spambots & Social spambots & Fake followers  \\
    \hline
    1 & Std. dev. of adjective frequency & Tweet sentiment arousal entropy  & Max. friend-follower ratio \\
    2 & Mean follower count & Mean friend count & Std. dev. of tweet inter-event time \\ 
    3 & Tweet content word entropy & Mean adjective frequency & Mean follower count\\
    4 & Max. friend-follower ratio & Min. favorite count & User tweet-retweet ratio \\
    5 & Max. retweet count & Tweet content word entropy &  Mean tweet sentiment happiness \\
    \hline
  \end{tabular}
  }
\end{table}

The second observation is that different bot classes have different sets of informative features.
Random Forest classifiers let us identify the most informative features to illustrate this point.
Table~\ref{tab:botclassFeatures} lists the five most informative features in classifiers trained on three different bot classes. 
We can see that traditional spambots generate a lot of content promoting products and can be detected by the frequent use of adjectives; social spambots tend to attack or support political candidates, therefore sentiment is an informative signal; finally, fake followers tend to have aggressive following patterns, flagged by the friend/follower ratio.

\begin{figure}
  \centering
  \includegraphics[width=0.8\linewidth]{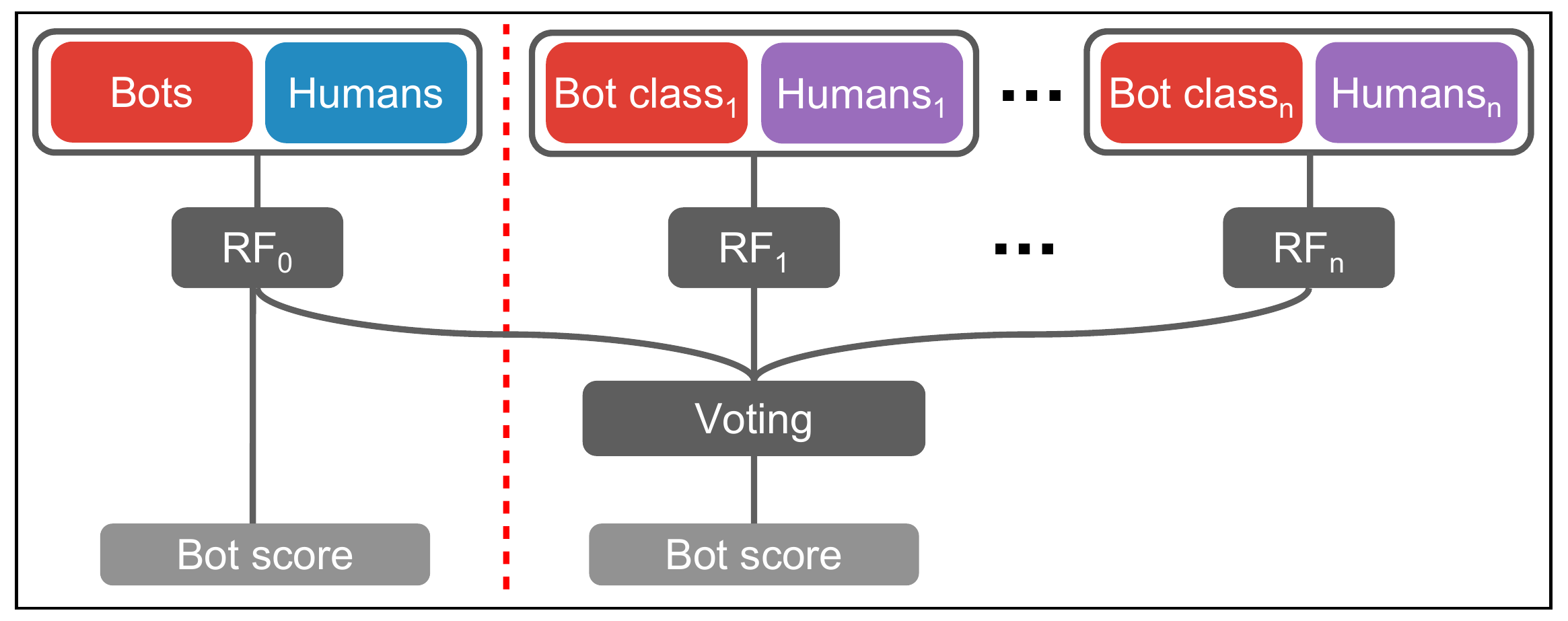}
  \caption{
  The architecture of the ESC model, used in Botometer V4. Source:  \protect\citeA{sayyadiharikandeh2020detection}.
  }
  \label{fig:esc_architecture}
\end{figure}

These observations lead to the Ensemble of Specialized Classifiers (ESC) architecture illustrated in Figure~\ref{fig:esc_architecture}.
In this framework, human accounts have a dedicated model $RF_0$, and each bot class also has a dedicated model ($RF_1 \dots RF_n$, respectively).
The final bot score is calculated by a voting scheme for the classifiers in the ensemble.
Among the specialized bot classifiers, the one that outputs the highest bot score $s_i$ is most likely to have recognized a bot of the corresponding class $i$.
For the human classifier, a low bot score $s_0$ is a strong signal of a human account.
Combined, the winning class is defined as  $i^* = \arg \max_i \{ s'_i \}$ where 
\[
s'_i = \left\{
 \begin{array}{ll}
 1-s_i  & \mbox{if } i = 0 \\
 s_i & \mbox{else.}
\end{array}
\right.
\] 
The ESC bot score is obtained by calibrating the score $s_{i^*}$ using Platt's scaling~\cite{niculescu2005predicting}. 

The ESC architecture improves the generalizability of the ensemble model compared to a single Random Forest model trained on all the datasets combined~\cite{sayyadiharikandeh2020detection}.  
Moreover, the bot class label $i^*$  helps interpret the score by revealing the class of the bot to which the target account is most similar. 
In Botometer V4, the sub-scores $s_i$ are reported together with the final score.

\subsection{BotometerLite}

When Botometer V4 was released, a new model, BotometerLite, was added to the Botometer family~\cite{yang_scalable_2020}.
BotometerLite was created to enable fast bot detection for large-scale datasets.
As mentioned earlier, the speed of bot detection methods is bounded by social media platforms' rate limits.
Botometer V4, for example, requires an account's 200 most recent tweets and recent mentions as input.
The API call has a limit of 43,200 accounts per API key per day.
Many computational social science studies using Twitter data need to analyze millions of accounts, a task that would take weeks or even months with these limits.

For scalability, BotometerLite gave up most of the contextual information and now relies on just user metadata.
This metadata is contained in the so-called \textit{user object} provided by Twitter API.
The rate limit for user object lookup is 8.6M accounts per API key per day. 
This is over 200 times the rate limit that bounds Botometer. 
Moreover, each tweet collected from Twitter has an embedded user object.
This brings two extra advantages.
First, once tweets are collected, no extra queries are needed for bot detection.
Second, while user lookup always reports the most recent user profile, the user object embedded in each tweet reflects the user profile at the moment when the tweet is collected.
This makes bot detection on archived historical data possible.

\begin{figure}
  \centering
  \includegraphics[width=0.8\linewidth]{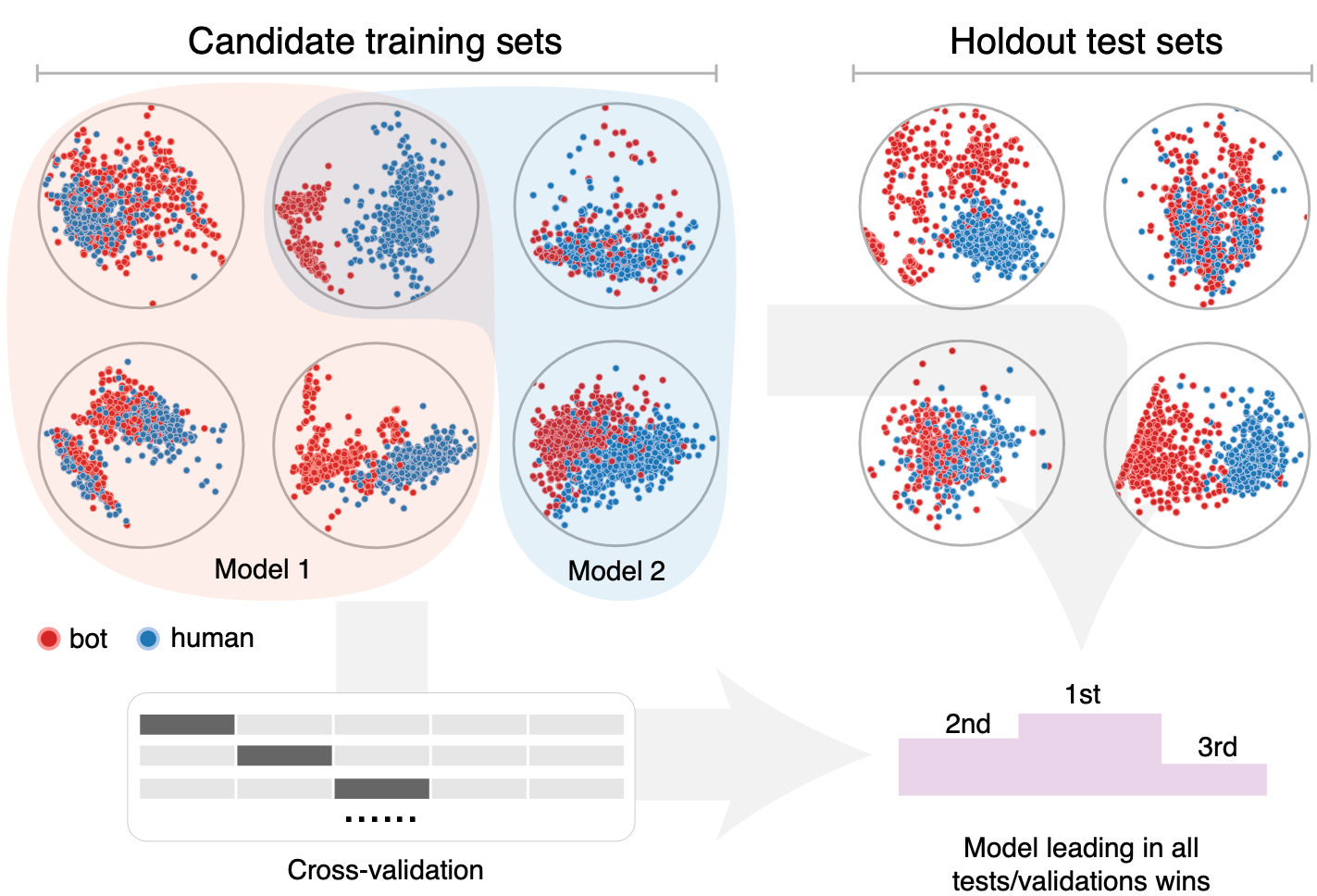}
  \caption{
  Illustration of the data selection mechanism for BotometerLite.
  The datasets from Table~\ref{table:dataset} are split into two groups: candidate training sets and holdout test sets.
  Random Forest classifiers are trained on different combinations of the candidate training sets.
  The winning classifier has to perform well in cross-validation on the training sets and on the holdout test sets.
  }
  \label{fig:botometerlite}
\end{figure}

In addition to improved scalability, BotometerLite employed a new data selection mechanism to ensure accuracy and generalizability.
Instead of throwing all training data into the classifier, the data selection mechanism aims to find a subset of training accounts that optimizes three evaluation metrics: cross-validation accuracy on the training data, generalization to holdout data, and consistency with Botometer (see Figure~\ref{fig:botometerlite}).
The data selection mechanism was inspired by the observation that some datasets might be contradictory to each other, as indicated by AUC scores below 0.5 in Figure~\ref{fig:generalizability}.
After evaluating the classifiers trained with all possible combinations of candidate training sets, the winning classifier only used five out of eight datasets but performed well in terms of all evaluation metrics.

Compared with Botometer, which tends to add more data and features, BotometerLite takes a different approach to simplifying the model by considering fewer features and datasets.
This leads to a compromise in accuracy and to the lack of bot class scores.
In return, BotometerLite lets researchers analyze large-volume streams of accounts in real time.

\section{A practical guide for bot detection}

In this Section, we provide a practical guide for computational social science practitioners who need to perform bot detection in their research.

Even though recent versions of Botometer used different methods to increase their generalizability to accounts outside of the training datasets, the challenge still exists.
For highly accurate bot detection, especially for accounts that are different from those in Table~\ref{table:dataset}, the best solution is still to annotate a batch of bot and human accounts in the target context.
Researchers can then apply existing bot detection frameworks to train classifiers on their own.
However, this might not be feasible in many cases.

The next solution is to find an existing bot detection method. 
Botometer is one of the best choices not only because it is carefully maintained and widely tested but also because it can be easily accessed.
The readers can test Botometer through its web interface.\footnote{\url{botometer.org}}
The only requirement is a Twitter account.
For programmatic access to the service, researchers can use the Botometer Pro API hosted by RapidAPI.com.\footnote{\url{rapidapi.com/OSoMe/api/botometer-pro}}
For API consumers, a valid Twitter developer account and a RapidAPI.com account are necessary.
Consumers have the freedom to use any programming language to query the API endpoints.
But the easiest way is through the official \texttt{botometer-python} library.\footnote{\url{github.com/IUNetSci/botometer-python}}
After installing the Python package, consumers can access the API with a few lines of code.
Examples can be found in the library's GitHub repository.

When analyzing the results collected for research, it is tempting to dichotomize the scores with an arbitrary threshold and consider accounts with scores above it bots.
This approach can be problematic because social bots may display a mixture of automatic and manual behaviors to avoid detection.
Instead, we believe it is more informative to inspect the distribution of scores over a sample of accounts.
For instance, when the goal is to compare the automation level of two account groups (e.g., a target group vs. a baseline group), researchers can employ statistical tests to compare the distribution directly.
In cases when binary classification is necessary, researchers may consider running analyses with different threshold choices to test the robustness of their findings.

\section{Conclusion}

In this Chapter, we provided an overview of research about social bots.
Our goal is to bring awareness about the need to properly handle social bots in computational social science research.
For readers interested in detection methods for social bots, we used Botometer as an example and provided a review of its recent developments, with a focus on generalizability and scalability issues.
We concluded with a practical guide for researchers who need to perform bot detection in their research.

With more people becoming aware of the existence and the impact of malicious social bots~\cite{pew2018social,varol2018deception}, social media platforms have implemented more aggressive measures to remove them.
However, malicious bots are unlikely to disappear for good.
Instead, they will evolve and rejoin the online community until new detection methods screen them out.
For instance, recent evidence suggested that bot operators started to employ state-of-the-art artificial intelligence tools to super-charge social bots~\cite{yang2023anatomy}, making their detection more challenging.
This long-lasting arms race calls for more efforts from computational social science research not only to provide technical solutions for detecting and handling evolving social bots but also to contemplate the philosophical questions raised by information ecosystems inhabited by bots.

\section{Further readings}

Readers interested in the technical details of Botometer can refer to the studies by \citeA{yang_arming_2019}, \citeA{sayyadiharikandeh2020detection}, and \citeA{yang_scalable_2020}.
\citeA{orabi_detection_2020} and \citeA{cresci2020decade} provide excellent reviews of other bot detection methods and tools.

\bibliographystyle{apacite}
\bibliography{ref}

\end{document}